\documentclass[journal, final, 10pt]{IEEEtran}

\ifCLASSINFOpdf
\else
   \usepackage[dvips]{graphicx}
\fi
\usepackage{url}

\usepackage{lipsum}
\usepackage{graphicx}
\usepackage{adjustbox}
\usepackage{graphicx, subcaption}
\usepackage{optidef}
\usepackage{algorithm, comment}
\usepackage{algorithmic}
\usepackage{amsmath,amsfonts,amssymb,amsthm, bm}
\usepackage{mathtools, enumitem}
\usepackage{tikz}
\usepackage{subcaption, float, comment} 
\usepackage{tikz-3dplot}
\usepackage{subcaption, float, comment} 
\usetikzlibrary{shapes,decorations}
\usetikzlibrary{positioning}

\newcommand\norm[1]{\left\lVert#1\right\rVert}
\DeclareMathAlphabet{\mathpzc}{OT1}{pzc}{m}{it}

\DeclareMathOperator*{\argmin}{arg\,min}

\newcommand{\G}{\mathcal{G}}

\newcommand{\Bx}{\mathbf{x}}
\newcommand{\By}{\mathbf{y}}

\newcommand{\Tr}{\textsf{T}}
\newcommand{\Tra}{\text{tr}}

\begin{document}
\title{Graph learning under spectral sparsity constraints}
%
\author{{B. Subbareddy$^{\ast \dag}$, Aditya Siripuram$^{\ast}$, Jingxin Zhang$^{\dag}$\\
$^{\ast}$ Indian Institute of Technology Hyderabad, India.\\
$^{\dag}$ Swinburne University of Technology, Australia.\\
}}
\maketitle
\begin{abstract}
Graph inference plays an essential role in machine learning, pattern recognition, and classification. Signal processing based approaches in literature generally assume some variational property of the observed data on the graph. We make a case for inferring graphs on which the observed data has high variation. We propose a signal processing based inference model that allows for wideband frequency variation in the data and propose an algorithm for graph inference. The proposed inference algorithm consists of two steps: 1) learning orthogonal eigenvectors of a graph from the data; 2) recovering the adjacency matrix of the graph topology from the given graph eigenvectors. The first step is solved by an iterative algorithm with a closed-form solution. In the second step, the adjacency matrix is inferred from the eigenvectors by solving a convex optimization problem. Numerical results on synthetic data show the proposed inference algorithm can effectively capture the meaningful graph topology from observed data under the wideband assumption.
\end{abstract}
\IEEEpeerreviewmaketitle

\IEEEPARstart{G}{raph} Signal Processing (GSP) provides a framework for processing data which is unstructured, complex, and massive. Such data is ubiquitous including the data from brain networks \cite{bullmore2009complex}, sensor works, gene networks \cite{ratnasamy1999inference} and transport networks  \cite{shuman2013emerging} as examples. GSP handles such complex data by effectively capturing the underlying relationship using graphs. In most of these domains, the signal of interest is more naturally indexed by the vertices of an underlying graph. The additional information from graph topology (which possibly encodes latent domain constraints) potentially allows better signal processing techniques than the classical framework. Research over the past decade has been successful in applying conventional signal processing techniques, like rate of change or frequency properties, to graphs \cite{ortega2018graph}. Among the many sub-fields in this research include designing graph filters, sampling \cite{8969404}, graph neural networks \cite{gama2018convolutional} and graph learning from data \cite{dong2018learning}. In the graph learning paradigm (the focus of this paper), the goal is to infer the underlying graph structure, given the data as signals (indexed by the vertices of such graph), with some assumptions on the relationship between the graph and the signal space. Most of these assumptions are based on the graph spectrum: this is indicative of how much a signal is allowed to change across an edge of the graph. Graph learning has found applications in diverse areas such as machine learning, biological network, and sensor networks \cite{dong2016learning}. The papers \cite{dong2018learning}, \cite{mateos2019connecting} have surveyed most of the work so far on graph learning from the signal processing perspective.

Graph reconstruction from the given data requires certain assumptions on how the data is related to the graph. Global smoothness based approaches essentially assume all observed signals have low graph frequencies (i.e., vary smoothly over the edges of the graph) \cite{kalofolias2016learn}, \cite{dong2018learning}, \cite{chepuri2017learning}. Filtering based techniques essentially assume a linear map from the unknown input to the observed signals, and hence try to force all of the observed data into the same graph frequency spectrum \cite{pasdeloup2017characterization}, \cite{segarra2016network}, \cite{mei2015signal}. Diffusion based approaches assume the data is generated as a smooth diffusion process starting from a few heat sources \cite{egilmez2018graph}, \cite{maretic2017graph}, \cite{thanou2017learning}, thus implicitly force a low pass filtering on the graph signal spectrum.


\begin{figure}[htbp]
\centering 
\hspace{-11mm}
\begin{subfigure}[b]{0.15\textwidth}
\centering
\begin{tikzpicture}[scale=0.9]
    \node[shape=circle,fill=red,draw=black, inner sep=1pt, outer sep = 0pt] (3) at (0,0) {3};
    \node[shape=circle,draw=black, inner sep=1pt, outer sep = 0pt] (1) at (0,1.5) {1};
    \node[shape=circle,draw=black, inner sep=1pt, outer sep = 0pt] (4) at (1.5,0) {4};
    \node[shape=circle,fill=red,draw=black, inner sep=1pt, outer sep = 0pt] (2) at (1.5,1.5) {2};

    \path [-] (3) edge node[left] {} (1);
    \path [-](1) edge node[left] {} (2);
    \path [-](4) edge node[left] {} (2);
    \path [-] (3) edge node[left] {} (4);
\end{tikzpicture}
\caption{}
\end{subfigure}
\hspace{-5mm}
\begin{subfigure}[b]{0.15\textwidth}
\centering 
\begin{tikzpicture}[scale=0.6]
    \node[shape=circle,draw=black, inner sep=1pt, outer sep = 0pt] (3) at (0,0) {3};
    \node[shape=circle,fill=red,draw=black, inner sep=1pt, outer sep = 0pt] (1) at (0,1.5) {1};
    \node[shape=circle,fill=red,draw=black, inner sep=1pt, outer sep = 0pt] (4) at (1.5,0) {4};
    \node[shape=circle,draw=black, inner sep=1pt, outer sep = 0pt] (2) at (1.5,1.5) {2};
    \node[shape=circle,fill=red,draw=black, inner sep=1pt, outer sep = 0pt] (5) at (3,1.5) {5};
    \node[shape=circle,draw=black, inner sep=1pt, outer sep = 0pt] (6) at (3,0) {6};
    \node[shape=circle,draw=black, inner sep=1pt, outer sep = 0pt] (8) at (1.5,-1.5) {8};
    \node[shape=circle,fill=red,draw=black, inner sep=1pt, outer sep = 0pt] (9) at (3,-1.5) {9};
    \node[shape=circle,fill=red,draw=black, inner sep=1pt, outer sep = 0pt] (7) at (0,-1.5) {7};

    \path [-] (3) edge node[left] {} (1);
    \path [-](2) edge node[left] {} (5);
    \path [-](4) edge node[left] {} (6);
    \path [-](1) edge node[left] {} (2);
    \path [-](4) edge node[left] {} (2);
    \path [-] (3) edge node[left] {} (4);
    \path [-] (3) edge node[left] {} (7);
    \path [-] (8) edge node[left] {} (4);
    \path [-] (6) edge node[left] {} (5);
    \path [-] (6) edge node[left] {} (9);
    \path [-] (8) edge node[left] {} (7);
    \path [-] (8) edge node[left] {} (9);

\end{tikzpicture}
\caption{}
\end{subfigure}
\begin{subfigure}[b]{0.15\textwidth}
\centering 
\begin{tikzpicture}[scale=0.7]
    \node[shape=circle,fill=red,draw=black, inner sep=1pt, outer sep = 0pt] (1) at (0,0) {1};
    \node[shape=circle,fill=red,draw=black, inner sep=1pt, outer sep = 0pt] (5) at (4,0) {5};
    \node[shape=circle,draw=black, inner sep=1pt, outer sep = 0pt] (2) at (1,1.5) {2};
    \node[shape=circle,draw=black, inner sep=1pt, outer sep = 0pt] (4) at (3,1.5) {4};
    \node[shape=circle,fill=red,draw=black, inner sep=1pt, outer sep = 0pt] (3) at (2,0) {3};

    \path [-] (1) edge node[left] {} (2);
    \path [-](2) edge node[left] {} (3);
    \path [-](4) edge node[left] {} (3);
    \path [-](5) edge node[left] {} (4);

\end{tikzpicture}
\caption{}
\end{subfigure}
\caption{Lateral inhibition system models}
\label{fig:lateral-graphs}
\end{figure}
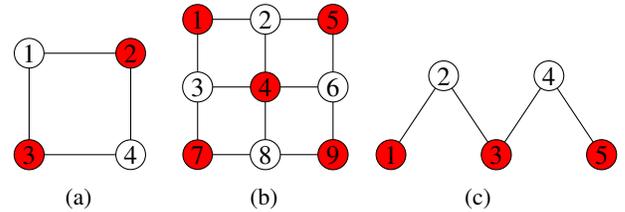

 

However, to the best of our knowledge, these approaches do not seem to allow the data to have a variety of graph frequencies. High frequency (negative correlation/high rate of change) across graph edges also conveys a structural relationship in the data. Hence it may not be prudent to allow only smooth variation. As discussed in \cite[pp-881]{huang2018graph}, \cite{huang2016graph}, one of the reasons GSP based analysis is appealing to brain imaging techniques is its ability to capture high frequency variation. For example, in  \cite{huang2016graph} the graph spectrum of brain signals during visual-motor learning tasks was decomposed into the low, band and high frequency components; and it was observed that most of the information of brain signals have both low and high graph frequencies. 

As an additional motivation for considering high frequencies, we consider lateral inhibition, a mechanism that occurs in neuro-biology and gene interactions \cite{sprinzak2010cis}, \cite{collier1996pattern}. Lateral inhibition is a cell to cell or neuron to neuron signal mechanism, where the excited neuron inhibits the action of the neighboring neurons. Figure \ref{fig:lateral-graphs} illustrates the process of lateral inhibition, where the red nodes are capacity excited neurons that reduce the activity of their neighbors (white nodes) \cite{ferreira2014pattern}. While the phenomenon of diffusion is more suitably captured by low frequencies, lateral inhibition is more suitably captured by high frequency variation on the graph. Existing methods do not seem to perform well when data has such high frequency variation (see the discussion in Section \ref{sec:results}).

Also, allowing the observed data to have high graph frequencies might result in a more compact and informative graph structural representation, which might be practically useful. This is the main motivation for our work in this paper.

To account for a wide range of frequencies, we propose a sparsity based graph learning model. Along similar lines to other works, we proceed in two phases - we first provide an algorithm for estimating the eigenvectors of the graph matrix and then proceed to find the eigenvalues from the eigenvectors in the second phase \cite{segarra2017network}. In principle, our model in this paper is similar to the block sparsity model of \cite{sardellitti2019graph}; however, the crucial difference is that we allow our data to have arbitrary frequency support including low-, high- and band-frequencies.



The paper is organized
as follows. In Section \ref{sec:GSP_intrduction}, we review graph signal processing along with notations used in the rest of the paper. In Section \ref{sec:graph_learning}, we introduce the notation, problem statement, and an overview of the proposed algorithm. In  section \ref{sec:algorithm} we give an overview of the proposed algorithm.  Subsequently, in section \ref{sec:results}, we discuss the results obtained using the proposed algorithm on synthetic data sets and provide a comparison with other well known graph learning algorithms. Finally, Section \ref{sec:con} concludes the work.

\section{Notations and problem formulation}
\label{sec:GSP_intrduction}

We assume the set of vertices of the underlying graph $\G$ is $V = \{1,2,\ldots,N\}$, and the set of edges is $E$, potentially with each edge having some non-negative weight. We denote by $A_\G$ the adjacency matrix of the graph $\G$:  an $N \times N$  matrix whose entries are the weights of the corresponding edges.  The degree matrix $D_\G$ is a diagonal $N \times N$ matrix whose diagonal entries are the number of edges from a given vertex (or, for weighted graphs, the sum of weights of edges incident on a vertex). We denote the Laplacian matrix of the graph $G$ as  $L_\G=D_\G-A_\G$. The subscript $\G$ is omitted if the graph is apparent from the context.

A graph signal $x \in \mathcal{R}^N$ is  a mapping $x: V \mapsto \mathbb{R}$. Similar to traditional signals, the notion of graph frequencies and graph Fourier transform of graph signals can be defined (\cite{shuman2013emerging}, \cite{sandryhaila2014big}, \cite{ortega2018graph}). For a graph signal $x$ defined on a graph $\G$, the Graph Fourier Transform (GFT) $y$ of $x$  is defined as $y = V_\G^{\Tr} x$ where  the $V_\G$ is the eigenvector matrix of  $A_\G$. The quadratic form $x^TL_\G x$ captures the \emph{variation} of the signal $x$ over the graph $L_\G$. Thus the eigenvectors of the Laplacian $L_\G$ have a natural rate of change interpretation: the eigenvectors corresponding to smaller eigenvalues do not vary much over the edges of the graph; and eigenvectors corresponding to larger eigenvalues have more variation. The number of nonzero coefficients in vector $y$ is $\norm{y}_0$.
In this work, we focus on adjacency-based graph transforms, though similar techniques may apply to Laplacian-based transforms as well.
\vspace{-0.7mm}
\subsection{Problem formulation}
\label{sec:graph_learning}
Given the set of observations $\Bx_1, \Bx_2, \ldots, \Bx_m \in \mathbb{R}^n$  from the unknown graph, the graph learning problem is to infer the unknown underlying graph. A popular graph learning algorithm \cite{dong2016learning} tries to find a graph  $\G$ such that the observations $\Bx_1, \Bx_2, \ldots, \Bx_m \in \mathbb{R}^n$ vary smoothly (or have low variation) on the graph:


	\begin{align} 
	 	\argmin_{\mathcal{L},Y} &\norm{X-Y}^{2}_{F}+\alpha Tr(Y^{\textsf{T}}\mathcal{L}Y)+\beta\norm{\mathcal{L}}^{2}_{F} \nonumber\\  &\text{s.t. }Tr(\mathcal{L}) =n, \mathcal{L}_{ij} =\mathcal{L}_{ji} \leq 0, ~~i\neq j, & \mathcal{L}.1=0 \nonumber
\end{align}

Here $X$ is the $N \times M$ observation matrix that has $\Bx_1, \Bx_2, \ldots, \Bx_m$ as its columns. The Laplacian quadratic form in the objective function encourages the observations to vary smoothly on the learned graph \cite{dong2016learning}. The Frobenius norm of $\mathcal{L}$ influences the number of edges of the learned graph.

As motivated in the introduction, we consider the scenario when the observed signals are not necessarily smooth on the underlying graph. We assume that the observed signals are a sparse combination of graph eigenvectors $V_{\G}$, as described in \eqref{eq:1}. Such a model allows for the underlying graph representation to be compact, especially in settings like lateral inhibition, where large variation on the graph is natural.

\begin{equation}
\label{eq:1}
\Bx_i = V_\G\ \By_i + \eta, \quad \norm{\By_i}_0 \leq k,
\end{equation}

A constraint in \eqref{eq:1} allows the observed data to be maximum $k$ linear combinations of the eigenvectors. Collecting all the observations $\By_i$'s into a matrix $Y$ gives the following model. 

\begin{equation}
  X=V_\G Y + \eta  
\end{equation}

 Here $Y$ is the GFT  coefficient matrix of data, and $\eta$ is the noise in observation.
Given $X$, mathematically, the problem of graph learning using spectral constraints is to solve the optimization problem \eqref{opt:2}: we try to minimize the error between the observed signal and the original signal $X$ by imposing the sparsity constraint on the observation matrix.

\begin{mini}
	{V_\G,Y}{\norm{X-V_\G\ Y}^{2}_{F}}{\label{opt:2}}{}
		\addConstraint{V_\G\ ^{\Tr}V_\G\ =I}{}
	\addConstraint{\norm{y_i}_0 \leq k, \quad \forall i \in \{1,\dots,m\}}{}
\end{mini}
 
 The first constraint (on $V_\G$) avoids the trivial solution and forces orthogonality on the eigenvector matrix. Solving \eqref{opt:2} is not straightforward as the objective function is non-convex due to the product of optimization variables. Moreover, the feasible set formed by the above constraints are non-convex due to the sparsity and orthogonality constraints. Our approach for finding the eigenbasis $V_\G$, coefficient matrix $Y$ and $\G$ are discussed in the following section. 

\section{Algorithms for graph learning under spectral sparsity constraints}
\label{sec:algorithm}
This section presents the algorithms developed for solving the optimization problem \eqref{opt:2}. We adopt the method of alternating minimization and tackle the sub-problems individually \cite{munkres1957algorithms}. We first fix the eigenbasis $V_\G$ and find the coefficient matrix $Y$, then fix the coefficient matrix $Y$ to find the eigenbasis $V_\G$. 

\subsection{Estimating the coefficient matrix $Y$}
\label{ssec:subhead}

Fixing the eigenbasis $V_\G$ in \eqref{opt:2} results in the convex objective function in \eqref{opt:3} with sparsity constraint, as in \eqref{opt:3}.

\begin{mini}[2]
	{Y}{\norm{X-V_\G\ Y}^{2}_{F}}{\label{opt:3}}{}
	\addConstraint{\norm{\By_i}_0 \leq k, \quad \forall i \in \{1,\dots,m\}}
\end{mini}

Though the constraint set in \eqref{opt:3} is non-convex, several relaxations can be used to obtain the convex problem. Since the matrix $V_\G$ is orthogonal, the solution is obtained by taking the top $k$ coefficients in any column of $V_\G^{\Tr}X$ \cite{elad2010sparse}. 

However, the sparsity $k$ may not be known in practice. We estimate the sparsity level $k$ by using the following technique: for each potential sparsity level $k$ (starting from $k=1$), we find $Y$ as described above, and compute the pseudo error
 \[
\frac{\norm{X-V_\G Y}_F^2}{\norm{X}_F^2}.
\]
We pick the value of $k$ for which the pseudo error is locally minimum. 

\subsection{Estimating the eigenbasis}

Once the coefficient matrix $Y$ is obtained from the previous step, we try to find the eigenbasis $V_\G$. We frame the following optimization problem  that requires finding the nearest orthogonal matrix to the observation data: 

\begin{mini}[2]
		{V_\G}{\norm{X-V_\G\ Y}^{2}_{F}}{\label{opt:4}}{}
	\addConstraint{V_\G\ ^{\Tr}V_\G\ =I}{}
\end{mini}

The optimization problem \eqref{opt:4} is an orthogonal Procrustes-problem \cite{gower2004procrustes}. The solution for such a problem is obtained by evaluating the singular value decomposition of matrix $X*Y^{\Tr}$. Thus a problem equivalent to the above is
\begin{equation}
\label{eq:trace-form}
    \max_{V_\G}  \Tra\left(Y^{\Tr}V_\G ^{\Tr}X\right).
\end{equation}
The solution for $V_\G$ is given by $V_\G=U_{1}U_{2}^{\Tr}$ where $U_1$ and $U_2$ are the singular vectors of  $X*Y^{\Tr}$: i.e. $X*Y^{\Tr} = U_1 \Sigma U_2^\Tr.$

The process is iterated until the convergence criterion is met. The algorithm is outlined below. 
 

\begin{algorithm}
    \caption{Finding eigenbasis vectors}
    \label{alg:altmin}
    \begin{algorithmic}
    \STATE \textbf{Given:} Obeservations $X$, sparse coefficients $k$
    \STATE \textbf{Initialisation:} Pick $V_\G $ random orthogonal matrix of $N \times N$

    \STATE $\text{currentObj} = \norm{X-V_\G\ Y}_{F}, \text{previousObj} = -\infty$ 
    \WHILE{$ \mid{\text{currentObj} - \text{previousObj}}\mid \leq \epsilon$}
    
    \STATE $Y \leftarrow V_\G^{\Tr}X $ to update Y
    \STATE $SVD(X Y^{\top}) =U_{1}\Sigma U_{2}^\Tr$
    
    \STATE $V_\G \leftarrow  U_{1}U_{2}^{\Tr} $ to update $V_\G$
    
    \STATE $\text{previousObj} \leftarrow \text{currentObj}$
    \STATE $\text{currentObj} \leftarrow \norm{X-V_\G\ Y}_{F}$
    \ENDWHILE
    \end{algorithmic}
    \end{algorithm}
    
Even though we are applying the standard co-ordinate descent framework, it is worthwhile to note that each of the resulting sub-problems  (sparse recovery problem and orthogonal Procrustes problem) is very well studied and has provable solutions. This probably explains the reasonably fast convergence of the proposed iterative technique.
 
\subsection{Estimating the adjacency matrix}

Once the eigenbasis $V_\G$ is known, the next step is to obtain the graph $\G$. To this end, we solve the constrained optimization problem \eqref{eq:adjacency_matrix} that enforces the properties of a valid adjacency matrix. This stage is similar to other such approaches in the graph learning literature \cite{segarra2017network}.


\begin{equation}
\label{eq:adjacency_matrix}
\begin{split}\Lambda = \argmin\{1 | A(\Lambda)_{ii}= 0, A(\Lambda) \geq 0, A(\Lambda)\bm{1} \geq \bm{1} \},
\end{split}
\end{equation}


  
\begin{figure}[htbp]
     \centering
     \begin{subfigure}[b]{0.2\textwidth}
         \centering
        \tdplotsetmaincoords{75}{110}
\begin{tikzpicture}[tdplot_main_coords,scale=0.25]
\tikzstyle{every node}=[font=\small]

\draw [dashed, tdplot_main_coords](0,0,0) -- (8,0,0) node[anchor=south]{};
\draw [dashed, tdplot_main_coords](0,0,0) -- (0,8,0) node[anchor=south]{};
\draw [dashed, tdplot_main_coords](0,8,0) -- (8,8,0) node[anchor=south]{};
\draw [dashed, tdplot_main_coords](8,0,0) -- (12,4,0) node[anchor=south]{};
\draw [dashed, tdplot_main_coords](8,8,0) -- (12,4,0) node[anchor=south]{};
\draw [dashed, tdplot_main_coords](0,8,0) -- (-4,4,0) node[anchor=south]{};
\draw [dashed, tdplot_main_coords](0,0,0) -- (8,8,0) node[anchor=south]{};

\draw [thin,draw=red,ultra thick, tdplot_main_coords](0,0,0) -- (0,0,-5) node[anchor=south]{};
\draw [thin,draw=blue,ultra thick, tdplot_main_coords](8,0,0) -- (8,0,1) node[anchor=south]{};
\draw [thin,draw=red,ultra thick, tdplot_main_coords](0,8,0) -- (0,8,-1) node[anchor=south]{};
\draw [thin,draw=blue,ultra thick, tdplot_main_coords](8,8,0) -- (8,8,5) node[anchor=south]{};
\draw [thin,draw=red,ultra thick, tdplot_main_coords](12,4,0) -- (12,4,-2) node[anchor=south]{};
\draw [thin,draw=blue,ultra thick, tdplot_main_coords](-4,4,0) -- (-4,4,2) node[anchor=south]{};

\draw [tdplot_main_coords, fill] (0,0,0) circle [radius=0.09] node [below] {3};;
\draw [tdplot_main_coords, fill] (8,0,0) circle [radius=0.09] node [below] {2};;
\draw [tdplot_main_coords, fill] (0,8,0) circle [radius=0.09] node [below] {5};;
\draw [tdplot_main_coords, fill] (8,8,0) circle [radius=0.09] node [below] {6};;
\draw [tdplot_main_coords, fill] (12,4,0) circle [radius=0.09] node [below] {1};;
\draw [tdplot_main_coords, fill] (-4,4,0) circle [radius=0.09] node [below] {4};;

\end{tikzpicture}
\caption{original graph}
       
         \label{fig:y equals x}
     \end{subfigure}
     \hfill
     \begin{subfigure}[b]{0.2\textwidth}
         \centering
         \tdplotsetmaincoords{75}{110}

\begin{tikzpicture}[tdplot_main_coords,scale=0.25]
\tikzstyle{every node}=[font=\small]

\draw [dashed, tdplot_main_coords](0,0,0) -- (8,0,0) node[anchor=south]{};
\draw [dashed, tdplot_main_coords](0,0,0) -- (0,8,0) node[anchor=south]{};
\draw [dashed, tdplot_main_coords](0,8,0) -- (8,8,0) node[anchor=south]{};
\draw [dashed, tdplot_main_coords](8,0,0) -- (12,4,0) node[anchor=south]{};
\draw [dashed, tdplot_main_coords](8,8,0) -- (12,4,0) node[anchor=south]{};
\draw [dashed, tdplot_main_coords](0,0,0) -- (-4,4,0) node[anchor=south]{};
\draw [dashed, tdplot_main_coords](0,8,0) -- (-4,4,0) node[anchor=south]{};
\draw [dashed, tdplot_main_coords](0,0,0) -- (8,8,0) node[anchor=south]{};

\draw [thin,draw=red,ultra thick, tdplot_main_coords](0,0,0) -- (0,0,-5) node[anchor=south]{};
\draw [thin,draw=blue,ultra thick, tdplot_main_coords](8,0,0) -- (8,0,1) node[anchor=south]{};
\draw [thin,draw=red,ultra thick, tdplot_main_coords](0,8,0) -- (0,8,-1) node[anchor=south]{};
\draw [thin,draw=blue,ultra thick, tdplot_main_coords](8,8,0) -- (8,8,5) node[anchor=south]{};
\draw [thin,draw=red,ultra thick, tdplot_main_coords](12,4,0) -- (12,4,-2) node[anchor=south]{};
\draw [thin,draw=blue,ultra thick, tdplot_main_coords](-4,4,0) -- (-4,4,2) node[anchor=south]{};

\draw [tdplot_main_coords, fill] (0,0,0) circle [radius=0.09] node [below] {3};;
\draw [tdplot_main_coords, fill] (8,0,0) circle [radius=0.09] node [below] {2};;
\draw [tdplot_main_coords, fill] (0,8,0) circle [radius=0.09] node [below] {5};;
\draw [tdplot_main_coords, fill] (8,8,0) circle [radius=0.09] node [below] {6};;
\draw [tdplot_main_coords, fill] (12,4,0) circle [radius=0.09] node [below] {1};;
\draw [tdplot_main_coords, fill] (-4,4,0) circle [radius=0.09] node [below] {4};;

\end{tikzpicture}
\caption{Our algorithm }
        
         \label{fig:three sin x}
     \end{subfigure}
     \hfill
     \begin{subfigure}[b]{0.2\textwidth}
         \centering
         \tdplotsetmaincoords{75}{110}

\begin{tikzpicture}[tdplot_main_coords,scale=0.25]
\tikzstyle{every node}=[font=\small]
\draw [dashed, tdplot_main_coords](-4,4,0) -- (8,0,0) node[anchor=south]{};
\draw [dashed, tdplot_main_coords](0,0,0) -- (0,8,0) node[anchor=south]{};
\draw [dashed, tdplot_main_coords](8,0,0) -- (0,8,0) node[anchor=south]{};
\draw [dashed, tdplot_main_coords](8,0,0) -- (8,8,0) node[anchor=south]{};
\draw [dashed, tdplot_main_coords](0,8,0) -- (8,8,0) node[anchor=south]{};
\draw [dashed, tdplot_main_coords](0,0,0) -- (12,4,0) node[anchor=south]{};
\draw [dashed, tdplot_main_coords](-4,4,0) -- (12,4,0) node[anchor=south]{};
\draw [dashed, tdplot_main_coords](0,8,0) -- (12,4,0) node[anchor=south]{};
\draw [dashed, tdplot_main_coords](0,0,0) -- (-4,4,0) node[anchor=south]{};
\draw [dashed, tdplot_main_coords](0,8,0) -- (-4,4,0) node[anchor=south]{};
\draw [dashed, tdplot_main_coords](-4,4,0) -- (8,8,0) node[anchor=south]{};

\draw [thin,draw=red,ultra thick, tdplot_main_coords](0,0,0) -- (0,0,-5) node[anchor=south]{};
\draw [thin,draw=blue,ultra thick, tdplot_main_coords](8,0,0) -- (8,0,1) node[anchor=south]{};
\draw [thin,draw=red,ultra thick, tdplot_main_coords](0,8,0) -- (0,8,-1) node[anchor=south]{};
\draw [thin,draw=blue,ultra thick, tdplot_main_coords](8,8,0) -- (8,8,5) node[anchor=south]{};
\draw [thin,draw=red,ultra thick, tdplot_main_coords](12,4,0) -- (12,4,-2) node[anchor=south]{};
\draw [thin,draw=blue,ultra thick, tdplot_main_coords](-4,4,0) -- (-4,4,2) node[anchor=south]{};

\draw [tdplot_main_coords, fill] (0,0,0) circle [radius=0.09] node [below] {3};;
\draw [tdplot_main_coords, fill] (8,0,0) circle [radius=0.09] node [below] {2};;
\draw [tdplot_main_coords, fill] (0,8,0) circle [radius=0.09] node [below] {5};;
\draw [tdplot_main_coords, fill] (8,8,0) circle [radius=0.09] node [below] {6};;
\draw [tdplot_main_coords, fill] (12,4,0) circle [radius=0.09] node [below] {1};;
\draw [tdplot_main_coords, fill] (-4,4,0) circle [radius=0.09] node [below] {4};;

\end{tikzpicture}
\caption{Global smoothness}
         \label{fig:five over x}
     \end{subfigure}
      \hfill
        \begin{subfigure}[b]{0.2\textwidth}
        \tdplotsetmaincoords{75}{110}
\centering
\begin{tikzpicture}[tdplot_main_coords,scale=0.25]
\tikzstyle{every node}=[font=\small]

\draw [dashed, tdplot_main_coords](0,0,0) -- (8,0,0) node[anchor=south]{};
\draw [dashed, tdplot_main_coords](0,8,0) -- (8,0,0) node[anchor=south]{};
\draw [dashed, tdplot_main_coords](8,8,0) -- (8,0,0) node[anchor=south]{};
\draw [dashed, tdplot_main_coords](0,0,0) -- (0,8,0) node[anchor=south]{};
\draw [dashed, tdplot_main_coords](0,8,0) -- (8,8,0) node[anchor=south]{};
\draw [dashed, tdplot_main_coords](0,0,0) -- (12,4,0) node[anchor=south]{};
\draw [dashed, tdplot_main_coords](0,0,0) -- (-4,4,0) node[anchor=south]{};
\draw [dashed, tdplot_main_coords](8,8,0) -- (-4,4,0) node[anchor=south]{};
\draw [dashed, tdplot_main_coords](0,0,0) -- (8,8,0) node[anchor=south]{};

\draw [thin,draw=red,ultra thick, tdplot_main_coords](0,0,0) -- (0,0,-5) node[anchor=south]{};
\draw [thin,draw=blue,ultra thick, tdplot_main_coords](8,0,0) -- (8,0,1) node[anchor=south]{};
\draw [thin,draw=red,ultra thick, tdplot_main_coords](0,8,0) -- (0,8,-1) node[anchor=south]{};
\draw [thin,draw=blue,ultra thick, tdplot_main_coords](8,8,0) -- (8,8,5) node[anchor=south]{};
\draw [thin,draw=red,ultra thick, tdplot_main_coords](12,4,0) -- (12,4,-2) node[anchor=south]{};
\draw [thin,draw=blue,ultra thick, tdplot_main_coords](-4,4,0) -- (-4,4,2) node[anchor=south]{};

\draw [tdplot_main_coords, fill] (0,0,0) circle [radius=0.09] node [below] {3};;
\draw [tdplot_main_coords, fill] (8,0,0) circle [radius=0.09] node [below] {2};;
\draw [tdplot_main_coords, fill] (0,8,0) circle [radius=0.09] node [below] {5};;
\draw [tdplot_main_coords, fill] (8,8,0) circle [radius=0.09] node [below] {6};;
\draw [tdplot_main_coords, fill] (12,4,0) circle [radius=0.09] node [below] {1};;
\draw [tdplot_main_coords, fill] (-4,4,0) circle [radius=0.09] node [below] {4};;

\end{tikzpicture}
\caption{Filter based approach}
\end{subfigure}

  \caption{Toy graph is used to motivate the proposed method with existing methods}
        \label{fig:toy_comparision}
\end{figure}
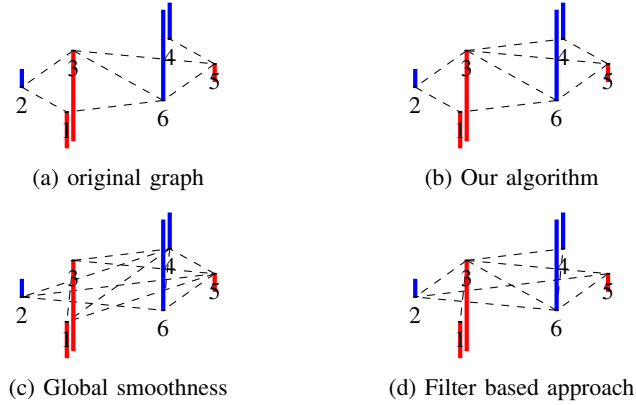

 To illustrate the advantage of proposed algorithm,  following exercise was performed by considering the graph with $N=6$ nodes and non-smooth graph signals (total variation is small). As discussed earlier, all state-of-the-art algorithms learn the graph assuming the graph signals are smooth. It is evident from Figure \ref{fig:toy_comparision} that, employing such methods to the graph signals with high total variation results in an inappropriate graph having a lot of false edges (edges that are not present in the original graph). Whereas employing our proposed algorithm results in a graph that is closer to the underlying graph with an added advantage of robustness against the small perturbations in the frequency spectrum. In the next section, we discuss the comparison methodology in detail.

\section{RESULTS AND DISCUSSION}
\label{sec:results}
We evaluate our algorithm on a synthetic data set. We first generate a ground truth graph and synthetic data set of graph signals for the ground truth graph. Then we apply our algorithm (and other algorithms in the literature) to compare the results. We use the following three models for generating the ground truth graph: 
\begin{itemize}
    \item RBF: Graphs are randomly generated with $N$ nodes (randomly in the unit square), and edges are placed based following procedure: first evaluate the distance between the nodes using distance metric as Euclidean distance), then assign edge weights by using $exp(-d(i,j)^{2}/2\sigma^{2})$ with  $\sigma= 0.5$, the function of the distance between the nodes. We only keep the edges with weights greater than $0.75$.
    \item ER: Erdos-Renyi model  \cite{erdHos1960evolution} with edge probability $0.2$ is used to create edges between the nodes with probability $0.2$.
    
    
     \item BA: Barabasi-Albert model  (a scale-free random graph) we build the graph adding one vertex at a time and using the preferential attachment mechanism \cite{barabasi1999emergence}. When adding a new vertex, we put edges from the new vertex with probability equal to the ratio of the degree of the existing vertex to the total number of existing edges in the graph.
    
    
\end{itemize}
 The ER and BA graphs in our experiments have unit edge weights. Similarly to other works \cite{segarra2017network}, \cite{dong2018learning}, \cite{kalofolias2016learn} we generate graphs with nodes $N=20$.
 
Starting with a ground truth graph, we generate $M=300$ graph signals according to our model in \eqref{eq:1}. Each of the graph signals is generated as a random sparse combination of eigenvectors of the graph previously generated, with the maximum sparsity $k_{max} = 5$. The non-zero coefficients are picked uniformly at random from $[1,2]$.
Before applying the learning algorithm, random Gaussian noise is added to the graph signals to create a synthetic data set. 

 The performance of the developed technique is evaluated with the widely used F-measure (harmonic mean of precision and recall) \cite{manning2010introduction} \cite{dong2018learning} \cite{kalofolias2016learn}. The F-measure calculates the deviation/similarity between the edges of the learned graph and those of the ground truth graph. The higher the F-measure is, the closer the learned and ground truth graphs are.
 
Once an algorithm is applied to the noisy synthetic data set, the F-measure is obtained. This is then averaged over $100$ independent realizations of noise, and further averaged over $100$ random graphs from a given model. The sizes of the graphs and the number of graphs over which averaging is done are similar to other graph learning algorithms \cite{dong2018learning}. Based on this framework, we perform the following experiments.

\subsubsection{Comparison with state-of-the-art algorithms}

The following table compares the F-measure of the proposed algorithm with other graph learning algorithms in the literature on the synthetic data set with noise level $0.3$. It is evident that the proposed algorithm outperforms the existing algorithms with significantly higher F-measure under the noise perturbation. This indicates that the proposed algorithm better captures the wideband graph frequency spectra than the existing algorithms.

\begin{center}
\begin{tabular}{ |c|c|c|c| } 
\hline
\multicolumn{4}{|c|}{F-measure comparison on noisy data set} \\
\hline
Algorithm & ER graph & BA graph & RBF graph  \\
\hline
Proposed & $\mathbf{0.8804}$ & $\mathbf{0.8964}$ & $\mathbf{0.9726}$ \\ 
Dong \cite{dong2018learning} & 0.3256 & 0.3138 & 0.4414 \\
Kalofolias \cite{kalofolias2016learn} & 0.3445 & 0.3260 & 0.4793\\
Segarra \cite{segarra2017network} & 0.2953 & 0.2787 &  0.4013 \\
Maretic \cite{maretic2017graph} & 0.2903 & 0.3089 & 0.4858\\
Chepuri \cite{chepuri2017learning} & 0.2117 & 0.2005 & 0.3172\\

\hline
\end{tabular}
\end{center}

\subsubsection{Effect of noise}
\begin{figure}[htb]
\centering
\includegraphics[width=7.5cm]{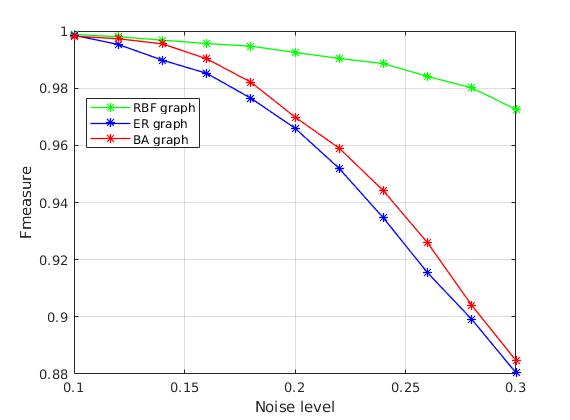}
\caption{ F-measure for different noise levels of different graphs. The noise level is variance of each entry of $\eta$ in \eqref{eq:1}}
\label{fig:noise}
\end{figure}

Figure \ref{fig:noise} plots the F-measure for our proposed algorithm as a function of the added noise level. It appears that our algorithm performs better for RBF than for ER and BA graphs. This is because ER and BA are unweighted graphs, and thus their inference is prone to discretization errors in our framework: the optimization problem in \eqref{eq:adjacency_matrix} does not impose binary constraints on the entries of $A_\G$.

 
\subsubsection{Effect of sparsity}
\begin{figure}[htb]
\centering
\includegraphics[width=7.5cm]{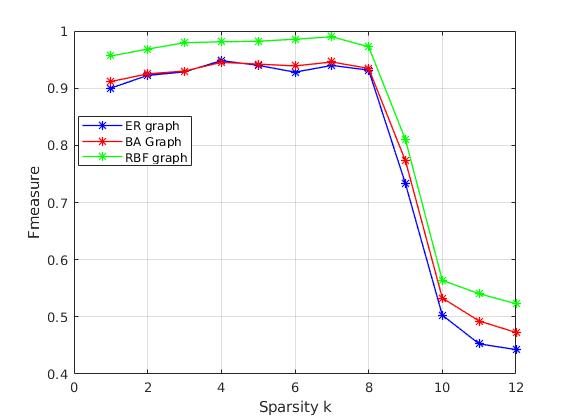}
\caption{ Plot of F-measure as a function of sparsity }
\label{fig:sparsity}
\end{figure}

Figure \ref{fig:sparsity} shows the impact of the spectrum sparsity of graph signals on the performance of the proposed algorithm. As shown, the F-measure decreases rapidly after about $40\%$ ($8$ out of $20$) of sparsity.

\section{CONCLUSIONS AND FUTURE WORK }
\label{sec:con}

We have proposed a novel model to learn graphs from wideband graph signal data, motivated from problems in neuro-imaging and lateral inhibition, an efficient algorithm for associating a graph topology to such data. The proposed method includes two steps in: 1) find a Fourier basis from observed signals by an iterative algorithm having closed form solution in each iteration; 2) infer the adjacency matrix from the estimated orthonormal basis by solving the optimization problem (\ref{eq:adjacency_matrix}). A comparison of our algorithm with existing graph learning algorithms under different graph signal spectra is presented to show the advantages of our method. Future research directions include testing the proposed algorithm on real data sets from neuro-imaging, further understanding the theoretical basis of the proposed algorithm and Laplacian based reconstruction with prior knowledge on one of the eigenvectors. 


\end{document}